\documentclass[a4paper]{jpconf}
\usepackage{graphicx}
\begin{document}
\title{Loophole in $K \to \pi \nu \bar\nu$ Search \\
 \ \& $K_L \to \pi^0 \nu \bar\nu$ Beyond Grossman--Nir Bound}

\author{George W.S. Hou}

\address{Department of Physics, National Taiwan University, Taipei 10617, Taiwan}

\ead{wshou@phys.ntu.edu.tw}

\begin{abstract}
The Grossman-Nir bound concept is robust, but the common perception of
$K_L \to \pi^0\nu\bar\nu < 1.4 \times 10^{-9}$ may be circumvented.
Because of the blinding $K\pi2$(3) decay, $K^+$ decay experiments
such as E787/E949 and the currently running NA62 kinematically
exclude the ``blinding zone''.
%There is a second exclusion region corresponding to background.
%
This offers an opportunity for the currently running KOTO experiment.
With no kinematic control, if the KOTO experiment is able to veto all
true background, it may discover $K_L \to \pi^0 + X^0$,
where $X^0$ is an invisible ``dark'' boson with mass $m_{X^0} \sim m_{\pi^0}$.
This could happen with 2015 data at hand,
which supposedly should allow KOTO to reach nominal GN bound sensitivity.
An explicit model of gauged $L_\mu - L_\tau$, linked with muon $g-2$
hence with very light $Z'$, is given.
Adding vector-like quarks that mix with usual quarks,
$W$-boson loops induce $s\to dZ'$ and $b\to sZ'$ transitions.
Besides realizing the scenario, the model provides further
illustration of potential impact.
Given that usual dark bosons linked with the photon are tightly constrained,
KOTO has a unique opportunity to probe the muon-related dark sector.
\end{abstract}

\section{The Kaon, as old as Particle Physics
}

If we count ``V particles" as the start, kaon physics is 69 years and counting,
coinciding also in time with Schwinger's calculation of $g - 2 = \alpha/\pi$,
the first ``penguin".
The present day Standard Model (SM) was pretty much built up by
learning from kaons, including the term ``penguin diagram''.
But with NA62 at CERN and KOTO at J-PARC, Japan pursuing the measurements of
$K^+ \to \pi^+ \nu \bar\nu$ and $K_L \to \pi^0 \nu \bar\nu$, respectively,
kaons remain at the New Physics (NP) frontier.

I felt frustrated at the FPCP2016 conference held at Caltech in June,
as the KOTO talk still referred to the Grossman-Nir (GN) bound~\cite{Grossman:1997sk}
as if it were a firm ceiling.
After relative quietness at ICHEP2016,
I was gratified to learn, early in this conference,
that the KOTO paper based on 2013 data (``100 hours'') had appeared, with title
\begin{quote}
\begin{center}
{\it A new search for the $K_L \to \pi^0 \nu \bar\nu$ and $K_L \to \pi^0 + X^0$ decays}~\cite{KOTO2013}
\end{center}
\end{quote}
that acknowledges the possibility of $K_L \to \pi^0 + X^0$ decay~\cite{FHK2015},
where $X^0$ is an invisible particle with $m_{X^0} \sim m_{\pi^0}$.
The new 90\% C.L. limit of ${\cal B}(K_L \to \pi^0 \nu \bar\nu) < 5.1 \times 10^{-8}$
is worse than the existing bound from E391A~\cite{Ahn:2009gb}
due to one observed event versus $0.34 \pm 0.16$ expected background.
The bound ${\cal B}(K_L \to \pi^0 + X^0) < 3.7 \times 10^{-8}$
fares better, indicating that KOTO finds the observed event
less consistent with this two-body possibility.

The GN bound comes in two forms:
\begin{eqnarray}
{\cal B}(K_L \to \pi^0\nu\bar\nu) & < & 4.3 \times {\cal B}(K^+ \to \pi^+\nu\bar\nu)
 \label{eq:GN} \\
 & < & 1.4 \times 10^{-9}. \ \ \ \ \ {\rm (``GN\ bound")}
 \label{eq:GNcommon}
\end{eqnarray}
The factor of 4.3 in Eq.~(\ref{eq:GN}) arises mostly from
$\tau_{K_L}/\tau_{K^+}$~\cite{Grossman:1997sk} and isospin, and is basically sound.
However, the commonly \emph{perceived} ``GN bound'' is the numerical value
given in Eq.~(\ref{eq:GNcommon}), which
follows by inserting the E787/E949 value~\cite{Artamonov:2009sz}
for ${\cal B}(K^+ \to \pi^+\nu\bar\nu)$.\footnote{
 Thus, if the upper bound for ${\cal B}(K^+ \to \pi^+\nu\bar\nu)$ drops,
 so would the number in Eq.~(\ref{eq:GNcommon}).}
%Paraphrasing Taylor Swift, the theme of this talk is to ``\emph{Shake it off!}'', i.e.
%
In this talk, we elucidate why the ``GN bound''
of Eq.~(\ref{eq:GNcommon}) is not fool-proof.
Since the 2015 data should~\cite{KOTO2013} improve sensitivity by a factor of 20 or more,
and should approach the ``GN bound'', contrary to common perception,
there is discovery potential with data at hand.

\section{Blinding [$K^+ \to \pi^+$]$\pi^0$:  Blessed are the Blind}

We, or at least I, have stared at the E949 plot so many times, that one becomes numb.
This final plot of E787/E949~\cite{Artamonov:2009sz} shows two signal boxes,
and in between the two, many red circles or blue triangles representing
the blinding $K^+ \to \pi^+\pi^0$ decay.
The $K\pi2$ branching ratio of 21\% is 100 million times or more stronger
than the target of $K^+ \to \pi^+ +$ {\it nothing}.
When faced with blinding light, one would turn the head away, close one's eyes,
and lift the arms to block it.
The action corresponding to this figure of speech is \emph{kinematic control}:
use $K^+$ and $\pi^+$ momenta to block out the blinding zone.
Though E787/E949 use stopped $K^+$ while NA62 exploits decay in-flight,
kinematic exclusion leads to rather similar signal boxes.

E949 was not unaware of an issue at hand, and pursued~\cite{Artamonov:2009sz} the
measurement of ${\cal B}(K^+ \to \pi^+X)$ where $X$ is an invisible particle.
Nevertheless, this blinding spot lead to a bound that is worse than
${\cal B}(K^+ \to \pi^+\nu\bar\nu)$ by a factor of 200 for $m_{X^0} \sim m_{\pi^0}$,
corresponding to an earlier, relatively unimpressive bound~\cite{Artamonov:2005cu}
of ${\cal B}(\pi^0 \to \nu\bar\nu) < 2.7 \times 10^{-7}$.
Thus, if Nature is so inclined to throw us a ``dark'' $X^0$ boson with $\pi^0$-like mass,
then E787/E949 and NA62 would be caught unaware, because of the use of kinematic control.
There is a second ``window'' due to $K\pi3$ decays.
But somehow, through numerous proposal reviews, the notion of
this ``loophole'' was not carried over to the $K_L$ side,
and the mindset was fixed on the pursuit of $K\to \pi\nu\nu$.

Our general observation~\cite{FHK2015} came as a surprise, as it is almost trivial:
``Blessed are the Blind."
It's been said that $K_L \to \pi^0 \nu \bar\nu$  measurement is like
searching for ``Nothing to Nothing'' (just $\gamma\gamma$).
Not only the incoming $K_L$ momentum is not known,
one detects just two $\gamma$s that are consistent with a $\pi^0$,
but there is no ID to give the true $\pi^0$ energy-momentum.
With such lack of kinematic control, the strategy is to veto everything.
But clearly one cannot veto WILPs (Weakly Interacting Light Particles).
Thus, if Nature throws in $X^0$s through the above two blinding spots,
KOTO could luck out!

Put simply: the KOTO experiment at J-PARC could discover $K_L \to \pi^0X^0$
above the ``GN bound'' of Eq.~(\ref{eq:GNcommon}).
Given that 2015 data is supposedly able to bring sensitivity down to this level,
this means discovery could start ``now'', at any value below the E391A limit~\cite{Ahn:2009gb}.

So, what you learned at ``school'' was not quite right.
But seeing this possibility came quite inadvertently for us,
for our starting point was top changing neutral current (tCNC)
search at the LHC, and this is a story of LHC-related work
leading to interesting kaon physics.
Let us therefore turn to a motivated model as an existence proof.

\section{A Motivated Model (existence proof)}

The explicit model is gauged $L_\mu - L_\tau$ related to muon $g - 2$,
plus extra vector-like quarks.

Gauging the difference between muon and tauon numbers is attractive
in several ways. The difference renders it anomaly free~\cite{XGHe}
with SM particle content,
while avoiding the electron (two other $L_e - L_i$ models) allows it to be less constrained,
and the coupling with muon connects it potentially with muon $g - 2$.
The U(1) is broken by a singlet scalar $\langle \Phi \rangle = v_\phi/\sqrt2$,
giving the $Z^\prime$ mass $m_{Z^\prime} = g^\prime v_\phi$,
where $g^\prime$ is the gauge coupling.
To link with quarks, one could add vector-like quarks $Q$, $D$, $U$
that couple to the $Z^\prime$ boson, where
 $Q$ stands for left- and right-handed doublets,
 while $D$ and $U$ are left-right singlets.
$Q$, $D$ and $U$ would have gauge invariant mass terms,
but these exotic quarks mix with SM quarks through Yukawa couplings
to the exotic $\phi$ field.

As stated,
our original interest~\cite{FHK2015} was rare $t\to cZ^\prime$ top decays,
and we took interest in the model of Altmannshofer \emph{et al.}~\cite{Altmannshofer:2014cfa} (AGPY).
It is well known that a heavy $Z^\prime$ boson could account for
the so-called $P_5^\prime$ anomaly uncovered by the LHCb experiment
in angular analysis of $B \to K^*\mu^+\mu^-$ decays,
where a 3.7$\sigma$ deviation from SM expectation is seen for 1 fb$^{-1}$ data.
While the discrepancy persists for 3 fb$^{-1}$ data,
the significance unfortunately did not improve~\cite{Aaij:2015oid},
hence this anomaly needs confirmation with more data at LHC Run 2.
Nevertheless, invoking the gauged $L_\mu - L_\tau$ force on the muon side,
AGPY constructed~\cite{Altmannshofer:2014cfa} the model as described above,
where $v_\Phi$ that generates $m_{Z^\prime}$ induces also
the desired $bsZ^\prime$ couplings at tree level
through $Q$ and $D$ quarks.
We took note~\cite{FHK2015,FHK} that the $U$ quark induces $t_R \to c_R Z^\prime$
transition (see the effective insertion inside the loop of Fig.~1)
that is not linked with the $P_5^\prime$ anomaly,
hence remain unconstrained.
For the gauged $L_\mu - L_\tau$ interactions,
$Z^\prime \to \mu^+\mu^-$ decay typically occurs with 1/3 branching fraction,
much higher than the $Z$ boson, hence the ATLAS and CMS experiments
should pursue $t\to cZ^\prime \to c\mu^+\mu^-$ search.

Having invoked gauged $L_\mu - L_\tau$, it would be great if the model
could explain $P_5^\prime$ and muon $g-2$ anomalies simultaneously.
But checking constraints, (probably) to their surprise,
AGPY discovered~\cite{Altmannshofer:2014pba} that a not so well known process,
called ``neutrino trident'' production, or $\nu_\mu N \to \nu_\mu \mu^+\mu^- N$,
constrains the $Z'$ to be rather light,
\begin{equation}
m_{Z'} < 400\ {\rm MeV}.
 \label{eq:lightZ'}
\end{equation}
This precludes the $Z^\prime$ to be simultaneously responsible for
the muon $g-2$ and $P_5^\prime$ anomalies, as a heavy $Z^\prime$ is implied for the latter.
But it is intriguing to think that \emph{New Physics} behind
the muon $g-2$ anomaly could arise from a \emph{light particle}!
As is well known from dark boson studies, this is possible only because
Eq.~(\ref{eq:lightZ'}) implies the gauge coupling $g' < 10^{-3}$,
far weaker than the weak coupling
(but the $Z^\prime$ is not Dark Matter, as it decays quite fast).
However, being lighter than the kaon inspired us to
study the impact of this $Z^\prime$ on kaon and $B$ physics.
Our original purpose was to see if this could rule out the model.

\begin{figure}[h]
\begin{center}
%\includegraphics[width=60mm,height=40mm]{plots/twidth.pdf}
%\vspace{30mm}
{
 \includegraphics[width=63mm]{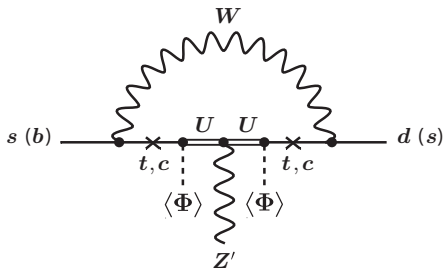}
}
\end{center}
%\vskip3.5cm
\vskip-0.15cm
\caption{
Effective $dsZ'$ ($sbZ'$) coupling, with $Z'$ coupled to a vector-like $U$ quark that
%mixes with $c$, $t$ (an ``$\times$'' flips chirality) and connect
mixes with $c$, $t$ (``$\times$'' flips chirality) and connect
with external $d$-type quarks via a $W$ boson loop.
}
\end{figure}

For the $Z^\prime$ related to muon $g-2$,
its lightness and the very precise measurements in
$B$ and $K$ sectors make it rather ``precarious''.
To avoid fine-tuning, one could decouple the $Q$ and $D$ quarks
by discrete $Z_2$ charge assignments.
But even with just the $U$ quark, as illustrated in Fig.~1,
a SM $W$-boson loop around the effective $ttZ^\prime$
($t$ can be interchanged with $c$) coupling turns it into~\cite{FHK2015, FHK}
$bsZ^\prime$ and $sdZ^\prime$ couplings, with help of chirality flip from $m_{t(c)}$.
The $sdZ^\prime$ effective coupling can precisely lead to $K \to \pi Z^\prime$
decay, where the light $Z^\prime$ is a candidate for the $X^0$
that could lead to the surprise evasion of the GN bound of
Eq.~(\ref{eq:GNcommon}), if $m_{Z^\prime} \sim m_{\pi^0}$.
It was through this model and by comparing $K^+$ and $K_L$ cases
that we stumbled upon the observation outlined in previous Section.
We note that chirality flip and CPV can both arise from SM,
and straightforward formulas can be found in Ref.~\cite{FHK},
including a more detailed discussion of $t\to cZ^\prime$
for both heavy (``$P_5^\prime$'') and very light (``muon $g-2$'') cases.

\section{What Then?  ---  an Illustration of Impact}

As KOTO analyzes their 2015 data,
what if they find $K_L \to \pi^0 +$ \emph{nothing} above $1.4 \times 10^{-9}$?
This is the nominal ``GN bound'' of Eq.~(\ref{eq:GNcommon}),
long perceived as impossible to breach,
but could actually happen with 2015 data.
We now go through possible hints and other possible rare $B$ and $K$ decays,
using the model presented in Sec. 3 as an illustration of what might lie ahead.
The discussion depends on the dimuon threshold.
Below this threshold, one has $Z^\prime \to \nu\bar\nu$ only,
while above it one could pursue $\mu^+\mu^-$ bump search.

It is interesting that the BaBar experiment has in fact a mild hint
for the analogous $B^+\to K^+\nu\bar\nu$ decay.
Note that ${\cal B}(B^+\to K^+\pi^0) \ll {\cal B}(K^+\to \pi^+\pi^0)$
is a rare decay itself at $10^{-5}$ level,
therefore there is no analogue of a $B^+\to K^+\pi^0$ induced ``blinding spot''.
Besides, the detection environment is rather different.
The BaBar experiment conducted~\cite{Lees:2013kla}
a binned $s_B \equiv m_{\nu\bar\nu}^2/m_B^2$ search for $B\to K^{(*)}\nu\bar\nu$,
separating $q^2 \equiv m_{\nu\bar\nu}^2$, or missing mass-squared, into 10 bins.
With 471M $B\bar B$ pairs, a \emph{two-sided} 90\% confidence interval was reported,
with the lower bound driven by a mild excess in the lowest $s_B$ bin.

For $m_{Z^\prime} = 135$ MeV, i.e. in the $\pi^0$ mass window,
hence $Z^\prime \to \nu\bar\nu$ 100\%,
we plot in Fig.~2[left] the BaBar allowed region in the $Y_{Ut}$--$Y_{Uc}$ plane,
where $Y_{Ui}$ (treated as real) are Yukawa couplings of $\Phi$ that mix the
exotic $U$ quark with quarks $i$ in SM.
We compare this with the
 E949 bound for $K^+ \to \pi^+ Z^\prime$,
 the $K_L \to \pi^0\nu\bar\nu$ search bound from E391a, and
 the commonly perceived ``GN bound'' of Eq.~(\ref{eq:GNcommon}).
We see that the BaBar result could by itself be
contentious with the latter,
and pushing below the E391a bound could possibly lead to discovery.

\begin{figure}[h]
\begin{center}
%\includegraphics[width=60mm,height=40mm]{plots/twidth.pdf}
%\vspace{30mm}
{
\begin{minipage}{2.7in}
 \includegraphics[width=2.65in]{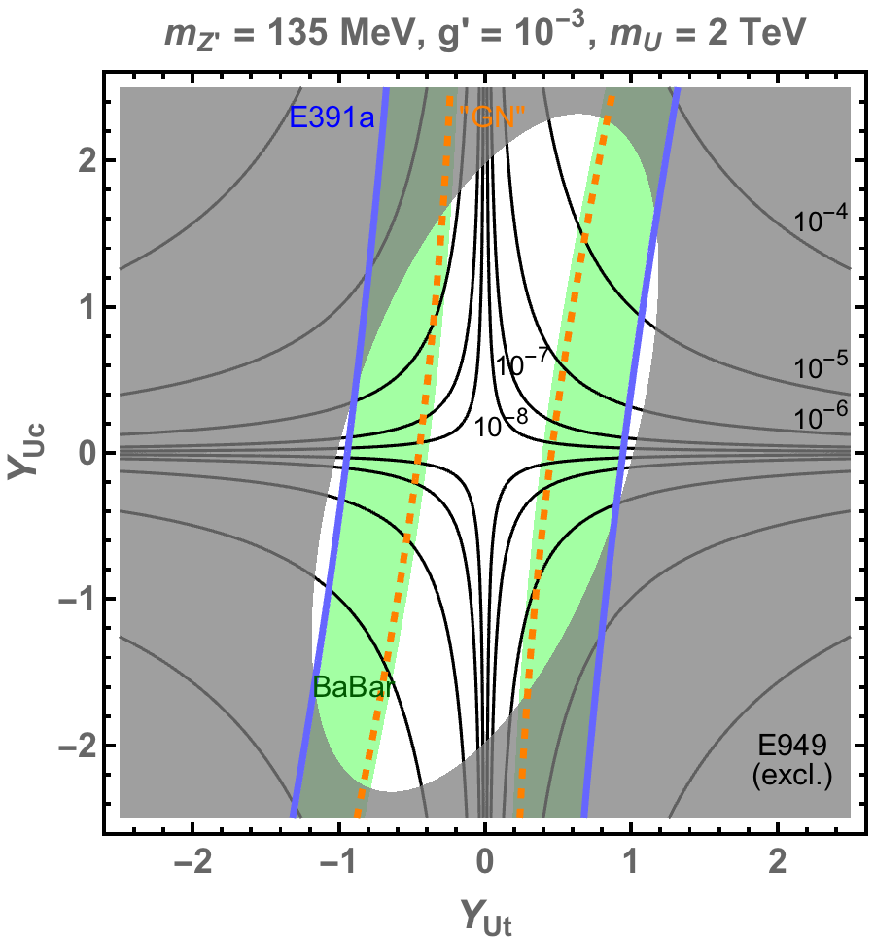}
\end{minipage}\hspace{2mm}%
\begin{minipage}{2.7in}
 \includegraphics[width=2.65in]{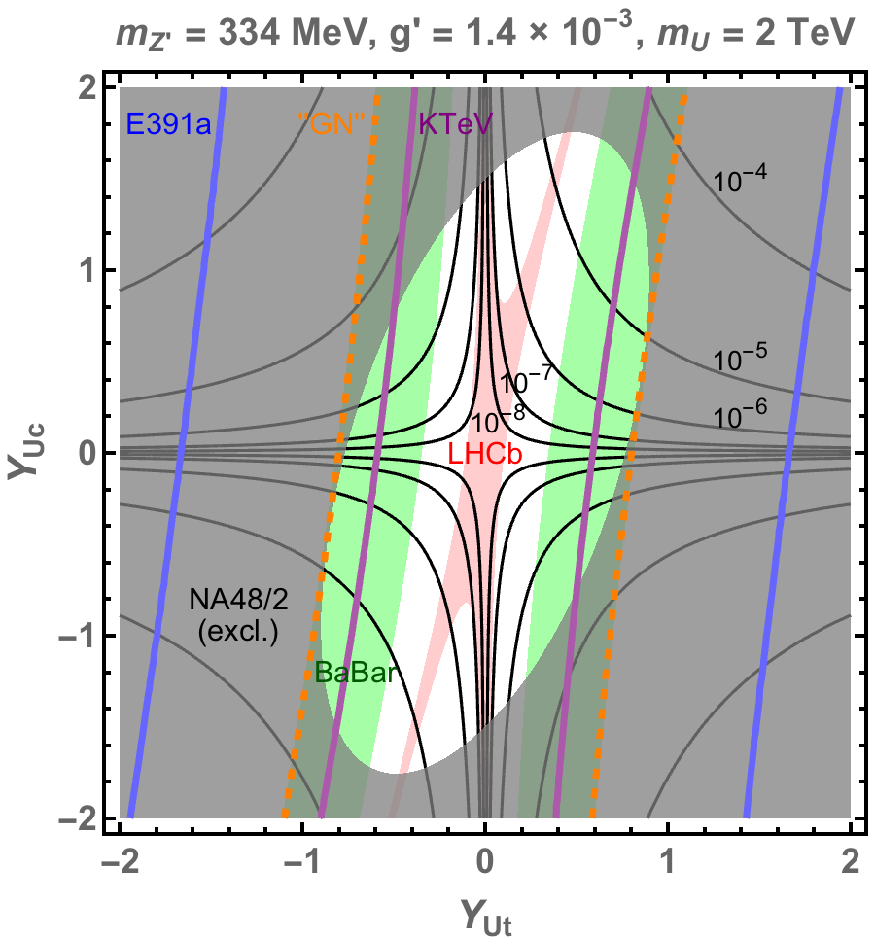}
\end{minipage}
}
\end{center}
%\vskip3.5cm
\vskip-0.15cm
\caption{
In $Y_{Uc}$-$Y_{Ut}$ plane,
for:
[Left] $m_{Z^\prime}=135$ MeV ($Z^\prime \to \nu\bar\nu$ at 100\%),
 with $\mathcal B(K^+\to\pi^+ Z^\prime) < 5.6\times 10^{-8}$ exclusion
 (E949, dark grey region);
[right] $m_{Z^\prime}=334$ MeV ($Z^\prime \to \mu^+\mu^-$ at 1/3),
 with
  95\% C.L. allowed by $B^0 \to K^{*0}\chi(\to \mu^+\mu^-)$
 (LHCb, pink region),
 $2\sigma$ exclusion by $K^+ \to \pi^+\mu^+\mu^-$
 (NA48/2, dark grey region), and
 90\% C.L. exclusion by $K_L \to \pi^0\mu^+\mu^−$
 (KTeV, solid purple line).
Common to both are the $2\sigma$ allowed range for
$\mathcal B(B^+\to  K^+Z^\prime)\mathcal B(Z^\prime\to\nu\bar\nu)
= (0.35^{+0.6}_{-0.15}) \times 10^{-5}$ (BaBar, light green region),
$\mathcal B(K_L \to \pi^0 Z^\prime) < 2.6\times 10^{-8}$ (E391a, blue solid line),
and the usual ``GN bound'' of
$\mathcal B(K_L\to \pi^0 Z^\prime) \mathcal B(Z^\prime \to \nu\bar\nu)
 < 1.4\times 10^{-9}$ (orange dashed line).
We plot $\mathcal B(t\to cZ^\prime)$ contours in the backdrop.
}
\end{figure}

So what about Belle?
Belle pioneered the $B^+\to K^+\nu\bar\nu$ search, but
its followup~\cite{Lutz:2013ftz} study just added 40\% data
without changing analysis strategy, where
a cut on high $p_{K^+}$ for sake of rejecting $B\to K^*\gamma$ background
precisely cuts against the $B\to K^{(*)}Z'$ possibility.
With existing full dataset, Belle should follow
the path as done by BaBar, i.e. the binned $m_{\rm mis.}^2$ analysis,
at least to crosscheck the BaBar result.
If some excess is also found, then the B factories
should attempt a combination.
This should be followed up at Belle II, whether
KOTO makes a discovery or not.
Competition should be welcome.

There is a second kinematic exclusion window for $K^+ \to \pi^+\nu\bar\nu$ search,
$m_{\rm mis.} > 260$ MeV, due to $K^+ \to \pi^+\pi\pi$ background.
In Ref.~\cite{FHK2015}, we had used the $m_{Z^\prime} = 285$ MeV case
to make estimates of what is allowed by a not so constraining
NA48/2 study~\cite{Batley:2011zz} of $K^+ \to \pi^+\mu^+\mu^-$,
and we alluded to~\cite{FHK2015} an apparent mild excess at lowest $q^2$
above the mean over $1.0\ {\rm GeV}^2 < q^2 < 6.0\ {\rm GeV}^2$
in the $B^0 \to K^{0}\mu^+\mu^-$ result~\cite{Aaij:2014pli} of LHCb.
Both cases have now been superseded, however, by the recent
dark boson bump search in $B^0 \to K^{*0}\mu^+\mu^-$ decay by LHCb~\cite{DarkLHCb},
and one is at best left with ``pocket'' physics, at the mercy of apparent
statistical fluctuations.
We give two examples, one near $m_{Z'} \sim 2m_\mu$ (below which efficiency vanishes),
another at $m_{Z'} \sim 334$ MeV, as can be read off the supplementary material to Ref.~\cite{DarkLHCb}.
Note that the $Z^\prime$ decays rather fast in the $L_\mu - L_\tau$ model,
even for this light range where the associated coupling $g^\prime \sim 10^{-3}$ is rather weak.
For the three cases we mention, the lifetimes are at sub-fs level~\cite{FHK},
hence one takes the $\tau = 0$~ps numbers from Ref.~\cite{DarkLHCb}.

For $m_{Z'} = 219$ MeV, just above the dimuon threshold,
though LHCb limit is weakest, the parameter space is still tightly
constrained by E949~\cite{Artamonov:2009sz}.
One finds $t \to cZ'$ must be less than $10^{-6}$ and typically much smaller,
and the case is overall less interesting.
Nevertheless, it should still be scrutinized by LHCb (and future Belle II)
with more data, as one should keep in mind the HyperCP dimuon events~\cite{Park:2005eka}.

Around $m_{Z'} = 334$ MeV, LHCb is again relatively tolerant,
and it is outside the signal box for E949 (and NA62).
We show the various constraints in Fig. 2[right].
The BaBar band (light green), now weaker than the ``GN bound'' of Eq.~(\ref{eq:GNcommon}),
is at some tension with the LHCb ``allowed'' region (pink),
while NA48/2~\cite{Batley:2011zz} result on $K^+ \to \pi^+\mu^+\mu^-$,
interpreted by us as a dimuon bump search,
replaces E949 as a constraint.
A search by KTeV for $K_L \to \pi^0\mu^+\mu^-$~\cite{AlaviHarati:2000hs}
provides a different constraint.
For this LHCb ``pocket'', $t \to cZ'$ could be enhanced
beyond $10^{-5}$ in the funnel region,
and would be rather interesting (if one believes in ``pockets'').
We have plotted the ${\cal}(t\to cZ')$ decay contours in the backdrop of Fig.~2.
More detailed discussion of $t\to cZ'$ decays can be found in Ref.~\cite{FHK},
including the $P_5^\prime$-motivated heavy $Z^\prime$ case.
In general, the rather low branching ratios would probably
need a 100 TeV $pp$ collider to probe.

%\footnote{
BaBar has recently searched~\cite{TheBABAR:2016rlg} for muonic dark force,
i.e. precisely the $L_\mu - L_\tau$ model,
by looking for a $Z^\prime$ emitted from $e^+e^- \to \mu^+\mu^-$
which subsequently decays via $Z^\prime \to \mu^+\mu^-$.
They ``exclude all but a sliver \ldots above dimuon threshold'',
i.e. some pockets are allowed below 400 MeV (and above as well).
%}
But the BaBar pockets do not seem to overlap with those of LHCb.
However, note that this, as well as the tension between LHCb
and the mild hint for $B \to K^{(*)} Z'(\to \nu\bar\nu)$
in Fig. 2[right] might be reconciled by enriching $Z'$ properties,
such as coupling to dark sector that changes the branching ratios.

Our highlight, especially in light of null results for
dimuon bump search from LHCb and BaBar,
is the $m_{Z'} \sim m_{\pi^0}$ window (Fig.~2[left]),
where opportunity lies for KOTO and Belle~(II), both with existing data,
and perhaps NA62, if the $\pi^0 \to$ \emph{nothing} bound~\cite{Artamonov:2005cu}
could be improved.
We remark that if KOTO reaches ``GN bound'' of Eq.~(\ref{eq:GNcommon})
without discovery, our point~\cite{FHK2015} remains valid.
That is, as NA62 (perhaps) improves the upper bound on $K^+ \to \pi^+\nu\bar\nu$,
KOTO could still make discovery in the $\pi^0$ mass window,
above the value implied by Eq.~(\ref{eq:GN}).

A very recent result by NA64~\cite{Banerjee:2016tad} claims
strong exclusion of sub-GeV dark photons. We offer a few remarks.
Dark ``photon'' invokes the idea of kinetic mixing with the photon,
hence coupling to electrons. In this context, note that a small region remains
near the $\pi^0$ mass window due to the ``blinding spot'' we discussed.
But in the model we use to illustrate, viz. gauged $L_\mu - L_\tau$,
the loop induced mixing of $Z'$ with photon is much below NA64
sensitivity. This is because NA64 is an electron beam-dump experiment.
A more challenging muon-beam experiment~\cite{Gninenko:2014pea} needs to be explored
to cover $L_\mu - L_\tau$ type of dark bosons.

\section{Conclusion}

$K_L \to \pi^0 +$ \emph{nothing} can occur above ``Grossman-Nir Bound''
of $1.4 \times 10^{-9}$, and the KOTO experiment is in the position
to make possible discovery with 2015 data at hand.
If KOTO sees early events, of course they should dutifully try hard
to reject them; but don't overkill --- don't kill the baby.
If one firmly finds events above the ``Grossman-Nir Bound'',
then likely one has a dark object of $\pi^0$ mass,
which could thereby slip through NA62 unnoticed.
Even when KOTO reaches below $1.4 \times 10^{-9}$,
the concept is still fertile, and one should not assume
the new GN bound derived from $K^+\to\pi^+\nu\bar\nu$ upper limits.
This should keep KOTO and Belle (II) in the game for a while,
and NA62 should try to improve the limit on $\pi^0 \to$ \emph{nothing}.

{\it ``Blessed are the Blind,
                            for they shall see the Heavens open.''}
surely sounds like the 9th Beatitudes that Jesus taught. Well, this is
not Scriptural, but all the same.
The ``blinding'' bright light may be emanating from an Angel,
or, as it is said ``even the Devil may masquerades himself
as an Angel of Light ...''.
But, no matter, for us mortals, we are just dying to see some
new ``light'' from Heaven, and the Devil is nothing but a fallen Angel,
which still offers a link.

So, please: Analyze, KOTO, Analyze!

\ack
We thank K. Fuyuto and M. Kohda for collaborative work,
and P. Crivelli, Y.B. Hsiung, M.~Kohda and M.~Pospelov for discussions.

\section*{Reference}


\begin{thebibliography}{99}

%\cite{Grossman:1997sk}
\bibitem{Grossman:1997sk}
  Y.~Grossman and Y.~Nir,
  %``K(L) ---> pi0 neutrino anti-neutrino beyond the standard model,''
  Phys.\ Lett.\ B {\bf 398}, 163 (1997).
%  [hep-ph/9701313].

\bibitem{KOTO2013}
  J.K.~Ahn {\it et al.} [KOTO Collab.],
  %``A New Search for the $K_{L} \to \pi^0 \nu \overline{\nu}$ and $K_{L} \to \pi^{0} X^{0}$ decays,''
  arXiv:1609.03637 [hep-ex];
  %%CITATION = ARXIV:1609.03637;%%
  see also talk by K. Shiomi, this proceedings.

\bibitem{FHK2015}
  K.~Fuyuto, W.-S.~Hou, M.~Kohda,
  Phys.\ Rev.\ Lett.\  {\bf 114}, 171802 (2015).

%\cite{Ahn:2009gb}
\bibitem{Ahn:2009gb}
  J.K.~Ahn {\it et al.}  [E391a Collab.],
  %``Experimental study of the decay K0(L) ---> pi0 nu nu-bar,''
  Phys.\ Rev.\ D {\bf 81}, 072004 (2010).
%  [arXiv:0911.4789 [hep-ex]].

%\cite{Artamonov:2009sz}
\bibitem{Artamonov:2009sz}
  A.V.~Artamonov {\it et al.}  [E949 Collab.],
  %``Study of the decay K+ ---> pi+ nu anti-nu in the momentum region 140 < P(pi) < 199-MeV/c,''
  Phys.\ Rev.\ D {\bf 79}, 092004 (2009).
%  [arXiv:0903.0030 [hep-ex]].
  %%CITATION = ARXIV:0903.0030;%%
  %85 citations counted in INSPIRE as of 28 Nov 2014

%\cite{Artamonov:2005cu}
\bibitem{Artamonov:2005cu}
  A.V.~Artamonov {\it et al.}  [E949 Collab.],
  %``Upper limit on the branching ratio for the decay pi0 ---> nu anti-nu,''
  Phys.\ Rev.\ D {\bf 72}, 091102 (2005).
%  [hep-ex/0506028].
  %%CITATION = HEP-EX/0506028;%%
  %24 citations counted in INSPIRE as of 28 Nov 2014

\bibitem{XGHe}
  X.-G. He, G.C. Joshi, H. Lew, R.R. Volkas,
  Phys.\ Rev.\ D \textbf{43}, 22 (1991).

%\cite{Altmannshofer:2014cfa}
\bibitem{Altmannshofer:2014cfa}
  W.~Altmannshofer, S.~Gori, M.~Pospelov, I.~Yavin,
  %``Dressing L_mu - L_tau in Color,''
  Phys.\ Rev.\ D {\bf 89}, 095033 (2014).
 % [arXiv:1403.1269 [hep-ph]].
  %%CITATION = ARXIV:1403.1269;%%
  %15 citations counted in INSPIRE as of 28 Nov 2014

%
\bibitem{Aaij:2015oid}
  R.~Aaij {\it et al.} [LHCb Collab.],
  %``Angular analysis of the $B^{0} \to K^{*0} \mu^{+} \mu^{-}$ decay using 3 fb$^{-1}$ of integrated luminosity,''
  JHEP {\bf 1602} (2016) 104.
 % doi:10.1007/JHEP02(2016)104
 % [arXiv:1512.04442 [hep-ex]].
  %%CITATION = doi:10.1007/JHEP02(2016)104;%%
  %57 citations counted in INSPIRE as of 26 Oct 2016

\bibitem{FHK}
  K.~Fuyuto, W.-S.~Hou and M.~Kohda,
  %``Z′ -induced FCNC decays of top, beauty, and strange quarks,''
  Phys.\ Rev.\ D {\bf 93}, 054021 (2016).
 % doi:10.1103/PhysRevD.93.054021
 % [arXiv:1512.09026 [hep-ph]].
  %%CITATION = doi:10.1103/PhysRevD.93.054021;%%
  %5 citations counted in INSPIRE as of 24 Oct 2016

%\cite{Altmannshofer:2014pba}
\bibitem{Altmannshofer:2014pba}
  W.~Altmannshofer, S.~Gori, M.~Pospelov, I.~Yavin,
  %``Neutrino Trident Production: A Powerful Probe of New Physics with Neutrino Beams,''
  Phys.~Rev.\ Lett.\  {\bf 113}, 091801 (2014).
 % [arXiv:1406.2332 [hep-ph]].
  %%CITATION = ARXIV:1406.2332;%%
  %4 citations counted in INSPIRE as of 28 Nov 2014

%\cite{Lees:2013kla}
\bibitem{Lees:2013kla}
  J.P.~Lees {\it et al.}  [BaBar Collab.],
  %``Search for $B → K^{(*)}ν\overlineν$ and invisible quarkonium decays,''
  Phys.\ Rev.\ D {\bf 87}, 112005 (2013).
%  [arXiv:1303.7465 [hep-ex]].
  %%CITATION = ARXIV:1303.7465;%%
  %12 citations counted in INSPIRE as of 29 Nov 2014

%\cite{Lutz:2013ftz}
\bibitem{Lutz:2013ftz}
  O.~Lutz, S.~Neubauer, M.~Heck, T.~Kuhr, A.~Zupanc {\it et al.}  [Belle Collab.],
  %``Search for $B \to h^{(*)} \nu \bar{\nu}$ with the full Belle $\Upsilon(4S)$ data sample,''
  Phys.\ Rev.\ D {\bf 87}, 111103 (2013).
%  [arXiv:1303.3719 [hep-ex]].
  %%CITATION = ARXIV:1303.3719;%%
  %9 citations counted in INSPIRE as of 29 Nov 2014

%
\bibitem{Batley:2011zz}
  J.R.~Batley {\it et al.}  [NA48/2 Collab.],
  %``New measurement of the K+- --> pi+-mu+mu- decay,''
  Phys.\ Lett.\ B {\bf 697}, 107 (2011).
%  [arXiv:1011.4817 [hep-ex]].
  %%CITATION = ARXIV:1011.4817;%%
  %23 citations counted in INSPIRE as of 28 Nov 2014
%
\bibitem{Aaij:2014pli}
  R.~Aaij {\it et al.}  [LHCb Collab.],
  %``Differential branching fractions and isospin asymmetries of $B \to K^{(*)} \mu^+ \mu^-$ decays,''
  JHEP {\bf 1406}, 133 (2014).
%  [arXiv:1403.8044 [hep-ex]].
  %%CITATION = ARXIV:1403.8044;%%
  %17 citations counted in INSPIRE as of 29 Nov 2014

\bibitem{DarkLHCb}
  R.~Aaij {\it et al.} [LHCb Collab.],
  %``Search for hidden-sector bosons in $B^0 \!\to K^{*0}\mu^+\mu^-$ decays,''
  Phys.\ Rev.\ Lett.\  {\bf 115}, 161802 (2015).
 % doi:10.1103/PhysRevLett.115.161802
 % [arXiv:1508.04094 [hep-ex]].
  %%CITATION = doi:10.1103/PhysRevLett.115.161802;%%
  %16 citations counted in INSPIRE as of 23 Oct 2016
%
\bibitem{Park:2005eka}
  H.~Park {\it et al.} [HyperCP Collab.],
  %``Evidence for the decay Sigma+ ---> p mu+ mu-,''
  Phys.\ Rev.\ Lett.\  {\bf 94}, 021801 (2005).
 % doi:10.1103/PhysRevLett.94.021801
 % [hep-ex/0501014].
  %%CITATION = doi:10.1103/PhysRevLett.94.021801;%%
  %97 citations counted in INSPIRE as of 24 Oct 2016

%
\bibitem{AlaviHarati:2000hs}
  A.~Alavi-Harati {\it et al.} [KTEV Collab.],
  %``Search for the Decay $K_L \to \pi^0 \mu^+ \mu^-$,''
  Phys.\ Rev.\ Lett.\  {\bf 84}, 5279 (2000).
 % doi:10.1103/PhysRevLett.84.5279
 % [hep-ex/0001006].
  %%CITATION = doi:10.1103/PhysRevLett.84.5279;%%
  %113 citations counted in INSPIRE as of 26 Oct 2016
%
\bibitem{TheBABAR:2016rlg}
  J.P.~Lees {\it et al.} [BaBar Collab.],
  %``Search for a muonic dark force at BABAR,''
  Phys.\ Rev.\ D {\bf 94}, 011102 (2016).
 % doi:10.1103/PhysRevD.94.011102
 % [arXiv:1606.03501 [hep-ex]].
  %%CITATION = doi:10.1103/PhysRevD.94.011102;%%
  %4 citations counted in INSPIRE as of 24 Oct 2016
%
\bibitem{Banerjee:2016tad}
  D.~Banerjee {\it et al.} [NA64 Collab.],
  %``Search for invisible decays of sub-GeV dark photons in missing-energy events at the CERN SPS,''
  arXiv:1610.02988 [hep-ex].
  %%CITATION = ARXIV:1610.02988;%%

%\cite{Gninenko:2014pea}
\bibitem{Gninenko:2014pea}
  S.N.~Gninenko, N.V.~Krasnikov and V.A.~Matveev,
  %``Muon g-2 and searches for a new leptophobic sub-GeV dark boson in a missing-energy experiment at CERN,''
  Phys.\ Rev.\ D {\bf 91}, 095015 (2015).
 % doi:10.1103/PhysRevD.91.095015
 % [arXiv:1412.1400 [hep-ph]].
  %%CITATION = doi:10.1103/PhysRevD.91.095015;%%
  %9 citations counted in INSPIRE as of 27 Oct 2016

\end{thebibliography}
\end{document}